\documentclass[twocolumn,aps,reprint,noshowpacs,superscriptaddress]{revtex4-2}

\usepackage{graphicx}% Include figure files
\usepackage{bm}
\usepackage{amssymb}
\usepackage{amsmath}
\usepackage{amsthm}
\usepackage{xcolor}
\usepackage{soul}

\usepackage[colorlinks=true,linkcolor=blue,urlcolor=blue,citecolor=blue,pdfauthor={ },pdftitle={ },pdfsubject={ },pdfkeywords={ }]{hyperref}
\graphicspath{{Figures/}}

\begin{document}
\title{Control-enhanced quantum metrology under Markovian noise}

\author{Yue Zhai}
\affiliation{Shenzhen Institute for Quantum Science and Engineering, Southern University of Science and Technology, Shenzhen, 518055, China}
\affiliation{International Quantum Academy, Shenzhen, 518055, China}
\affiliation{Guangdong Provincial Key Laboratory of Quantum Science and Engineering, Southern University of Science and Technology, Shenzhen, 518055, China}

\author{Xiaodong Yang}
\email{yangxd@sustech.edu.cn}
\affiliation{Shenzhen Institute for Quantum Science and Engineering, Southern University of Science and Technology, Shenzhen, 518055, China}
\affiliation{International Quantum Academy, Shenzhen, 518055, China}
\affiliation{Guangdong Provincial Key Laboratory of Quantum Science and Engineering, Southern University of Science and Technology, Shenzhen, 518055, China}

\author{Kai Tang}
\affiliation{Shenzhen Institute for Quantum Science and Engineering, Southern University of Science and Technology, Shenzhen, 518055, China}
\affiliation{International Quantum Academy, Shenzhen, 518055, China}
\affiliation{Guangdong Provincial Key Laboratory of Quantum Science and Engineering, Southern University of Science and Technology, Shenzhen, 518055, China}

\author{Xinyue Long}
\affiliation{Department of Physics, Southern University of Science and Technology, Shenzhen, China}
\affiliation{Shenzhen Institute for Quantum Science and Engineering, Southern University of Science and Technology, Shenzhen, 518055, China}
\affiliation{Guangdong Provincial Key Laboratory of Quantum Science and Engineering, Southern University of Science and Technology, Shenzhen, 518055, China}

\author{Xinfang Nie}
\affiliation{Department of Physics, Southern University of Science and Technology, Shenzhen, China}
\affiliation{Shenzhen Institute for Quantum Science and Engineering, Southern University of Science and Technology, Shenzhen, 518055, China}
\affiliation{Guangdong Provincial Key Laboratory of Quantum Science and Engineering, Southern University of Science and Technology, Shenzhen, 518055, China}

\author{Tao Xin}
\affiliation{Shenzhen Institute for Quantum Science and Engineering, Southern University of Science and Technology, Shenzhen, 518055, China}
\affiliation{International Quantum Academy, Shenzhen, 518055, China}
\affiliation{Guangdong Provincial Key Laboratory of Quantum Science and Engineering, Southern University of Science and Technology, Shenzhen, 518055, China}

\author{Dawei Lu}
\affiliation{Department of Physics, Southern University of Science and Technology, Shenzhen, China}
\affiliation{Shenzhen Institute for Quantum Science and Engineering, Southern University of Science and Technology, Shenzhen, 518055, China}
\affiliation{International Quantum Academy, Shenzhen, 518055, China}
\affiliation{Guangdong Provincial Key Laboratory of Quantum Science and Engineering, Southern University of Science and Technology, Shenzhen, 518055, China}

\author{Jun Li}
\email{lij3@sustech.edu.cn}
\affiliation{Shenzhen Institute for Quantum Science and Engineering, Southern University of Science and Technology, Shenzhen, 518055, China}
\affiliation{International Quantum Academy, Shenzhen, 518055, China}
\affiliation{Guangdong Provincial Key Laboratory of Quantum Science and Engineering, Southern University of Science and Technology, Shenzhen, 518055, China}

\begin{abstract}
Quantum metrology is supposed to significantly improve the precision of parameter estimation   by utilizing suitable quantum resources. However, the predicted precision can be severely distorted by realistic noises.
 Here, we propose a control-enhanced quantum metrology scheme to defend against these noises to improve the metrology performance. Our scheme can automatically alter the parameter-encoding dynamics with adjustable controls, thus leading to optimal resultant states that are less sensitive to the noises under consideration. 
As a demonstration, we numerically apply it to the problem of frequency estimation under several typical Markovian noise channels. By comparing our control-enhanced scheme with the standard scheme and the ancilla-assisted scheme, 
    we show that our scheme performs better and can
    improve the estimation precision up to around
    one order of magnitude.
Furthermore, we conduct a proof-of-principle experiment in a nuclear magnetic resonance system  to verify the effectiveness of the proposed scheme.  The research here is helpful for current quantum platforms to harness the power of quantum metrology in realistic noise environments.

\end{abstract}
\maketitle

\section{Introduction}
Quantum metrology concerns how to manipulate available quantum resources to acquire the best estimation precision of the parameters to be measured \cite{GI11,GT14,PS18}. The standard procedure of quantum metrology consists of preparing an input probe state, having it interacts with the encoding dynamics, and measuring the output state to extract the parameter. Ideally, in the absence of noise, quantum systems undergo unitary evolutions.  Additionally, it has been well established that entangled probe states with optimal measurements can achieve a precision improvement of the estimated parameter over  classical strategies, up to a factor of $1/\sqrt{N}$ in the number of particles $N$ \cite{GV06}. However, for noisy processes, the inevitable interplay with environments leads to nonunitary evolutions and limits the usefulness of the aforementioned quantum strategies \cite{ED11,EDR11}. For example, maximally entangled states can lose their advantages when subjected to dephasing effects \cite{HM97,FA17}. Under different noisy environments, the ultimate precision bounds that can be attained have attracted many theoretical studies \cite{KJ13,UF17,YS17,AM14,DR12}. Nevertheless, how to use available resources in realistic experiments to saturate these bounds is still an urgent research area \cite{UD01,CB13,BC15,AS16,ZB19}.

The harmfulness of these unavoidable noises can be mitigated by applying additional controls in the metrology process. 
Representative approaches include  dynamical decoupling \cite{TQH13,LJL15,PMW16} and quantum error correction \cite{WMF14,KLS14}, yet they are usually designed for special cases or need abundant extra resources and thus are rarely explored experimentally.
Carefully derived feedback controls have been proven to be helpful for maintaining the precision limit \cite{YF15,YH16} but are hard to obtain in experiments. Ancilla-assisted approaches \cite{DF14,HM16,WW18,HM18}, which utilize the entanglement and joint measurement of the system and the ancillary qubits, are certainly effective, but at the expense of involving extra qubits and measurements. 
Recently, researchers attempted to iteratively find optimal controls that interact with the encoding dynamics to improve the metrology performance through the use of a gradient-based algorithm  \cite{LJY17,PhysRevA.96.042114,yang2021hybrid,PhysRevResearch.2.033396,Qin_2022} or reinforcement learning algorithm \cite{HX19,xiao2022parameter}. These control-enhanced methods are general and flexible,  but the optimizations need hard-to-obtain gradient information  or  extensive training data, which restrict their experimental applications \cite{AS15,liu2022optimal}. Furthermore, the comparison with the other methods mentioned is still less explored.

In this work, we propose a control-enhanced quantum metrology scheme to tackle the noise issue.  It functions by iteratively refreshing adjustable controls to alter the encoding dynamics, thus automatically driving the system to certain states that are more robust to the noises under consideration. 
We use a gradient-free Nelder-Mead simplex algorithm \cite{nelder1965simplex} to accomplish the learning process, which requires fewer experimental resources and is more likely to reach global optimum than the gradient-based algorithms. Other gradient-free algorithms \cite{mueller_2022,Rossignolo_2022,yang2019improved} are also applicable here.
To demonstrate its effectiveness, we test it in the case of frequency estimation under common Markovian noise environments. We also compare the proposed scheme  with the ancilla-assisted scheme \cite{DF14,HM16,WW18,HM18}, showing that the precision can be improved by almost an order of magnitude.  Furthermore, we experimentally verify the proposed scheme in a nuclear-magnetic-resonance system considering pure dephasing noise. 
The outline of this study is described as follows. First, we introduce the control-enhanced quantum metrology scheme in Sec. \ref{framework}. Second, we conduct numerical simulations considering several kinds of  Markovian noise channels in Sec. \ref{numerical simulations }. Next, we show the results of experimental verifications in Sec. \ref{experiment}. Finally, some conclusions and a discussion  are provided in Sec. \ref{conclusion}.

  \begin{figure}
	\centering
	\includegraphics[width=0.48\textwidth,height=0.30\textwidth]{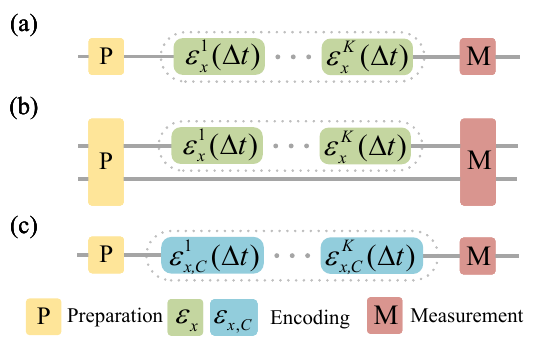}
	\caption{Comparison of three quantum metrology schemes. (a) Standard scheme. A theoretical optimal probe state $\rho_0$ (usually the maximally entangled state)  interacts with the encoding dynamics $\varepsilon_x$ over a time period $T$, which is divided into $K=T/\Delta t$ equal parts $\varepsilon_x^k(\Delta t)$, with $k=1,2,...,K$. Suitable measurements are then performed on the resultant state $\rho_x$ to extract the parameter information. (b) Ancilla-assisted scheme. The system and the ancillary qubit are first jointly prepared at the maximally entangled state, and then the system solely interacts with the sliced encoding dynamics $\varepsilon_x^k(\Delta t)$; finally, a joint measurement is performed. 
		(c) Control-enhanced scheme. The system is 
		started from an arbitrary initial probe state 
		$\rho_0$. The encoding dynamics is 
		engineered with adjustable controls, 
		marked as $
		\varepsilon_{x,C}^k(\Delta t)$ for each time
		length $\Delta t$. The resultant state $\rho_x$ is then 
		evaluated by  suitable measurements, and the 
		controls are iteratively refreshed by 
		a suitable optimization algorithm. This 
		procedure automatically alters the encoding 
		dynamics to engineer the initial probe to 
		some optimal one that is insensitive to the 
		noises under consideration. }
	\label{theory}
\end{figure}

\section{Framework}\label{framework}
Consider the task of estimating the parameter $x$ in the general form of Hamiltonian $\mathcal{H}_0(x)$ under specific Markovian noises.  The standard metrology scheme is to first prepare the system at some theoretically optimal probe state $\rho_0$, then have it interacts with the encoding dynamics $\varepsilon_x$, and finally perform suitable measurements, as shown in Fig. \ref{theory}(a). However, the metrology performance can be greatly affected by  noises, and determining how to manipulate available resources to achieve the best precision is challenging. 
It has been verified that the ancilla-assisted scheme \cite{DF14,HM16,WW18,HM18}, as shown in  Fig. \ref{theory}(b), moderately improves the metrology performance by entangling the system with the ancillary qubit but needs extra qubit resources and joint measurements.
Here, we propose a practical control-enhanced  scheme to alter the encoding dynamics such that the noises can be resisted for better metrology, as shown in  Fig. \ref{theory}(c).
With iteratively refreshed controls interacting with the encoding dynamics, the optimal resultant state can be automatically discovered, which is less sensitive to the noises. In the following, we describe our control-enhanced scheme in detail. 

\subsection{Simulation of the encoding dynamics}
We first describe how to simulate the encoding dynamics with controls and Markovian noises. 
The noiseless Hamiltonian of a controlled system can be written as
\begin{equation}\label{ham}
\mathcal{H}=\mathcal{H}_0(x)+\sum_{l=1}^L u_l(t) \mathcal{H}_l,
\end{equation}
where $\mathcal{H}_0(x)$ denotes the encoding Hamiltonian, $x$ is the parameter to be estimated,
and $u_l(t):t\in[0,T]$ represents the amplitude of the $l$th control field with respect to the control Hamiltonian $\mathcal{H}_l$.
Suppose that the system is in a Markovian environment; then the encoding dynamics can be described by the following Lindblad equation \cite{KP16,LD19} :
\begin{eqnarray} \label{linda}
\frac{d \rho}{d t}
&=& -i[\mathcal{H}, \rho]+\sum_v \gamma_v \left(L_v \rho L_v^{\dagger}-\frac{1}{2}\left\{L_v^{\dagger} L_v, \rho\right\}\right) \\ \nonumber
&\equiv & \mathcal{L}(\rho)\equiv -i\mathcal{H}^\times(\rho) +\Gamma (\rho),
\end{eqnarray}
where $\rho$ is the system state and $\mathcal{L}$, $\mathcal{H}^\times,$ and $\Gamma$ denote the superoperators for the total, noiseless, and noisy evolution dynamics, respectively. For the noisy part, the Lindblad operators $L_v$ are used to model various dissipative channels, and the constants $\gamma_v$ are the corresponding dissipative rates.

 To conveniently solve this equation, we use Liouville's representation to reshape the states and operators \cite{HF03}. By stacking the columns of the quantum state $\rho=\sum_{ij} \rho_{ij}|i\rangle \langle j|$, we get the corresponding vector representation, i.e., $\rho \rightarrow |\rho\rangle=\sum_{ij} \rho_{ij}|j\rangle \otimes |i\rangle$. Meanwhile, the effects of the operators $U$ and $V$ performed on $\rho$ can be rephrased as $U\rho V\rightarrow |U\rho V\rangle=V^T \otimes U|\rho\rangle$. In this way, the solution of the above equation can be formally written as 
\begin{equation}
	|\rho(t)\rangle= e^{\mathcal{\hat L}t}|\rho(0)\rangle, \mathcal{\hat L} =-i \hat{\mathcal{H}}^\times +\hat{\Gamma},
\end{equation}
where $\mathcal{\hat L}$, $\mathcal{\hat H}$, and $\hat{\Gamma}$ represent the superoperators in the Liouville representation. Precisely, $\mathcal{\hat H}$ and $\hat{\Gamma}$ satisfy the following transformation rules: $ \hat{\mathcal{H}}^\times =\mathbb{I} \otimes \mathcal{H}-\mathcal{H}^{*} \otimes \mathbb{I}$ and $\hat{\Gamma} =\sum_{v} \frac{\gamma_v}{2}( L_{v}^{*} \otimes L_{v}-\frac{1}{2} \mathbb{I} \otimes L_{v}^{\dagger} L_{v}-\frac{1}{2} L_{v}^{T} L_{v}^{*} \otimes \mathbb{I})$, where $\mathbb{I}$ represents the identity matrix.

\subsection{Evaluation of the metrology performance}
We proceed to describe how to evaluate the metrology performance by estimating $x$. We denote the system's initial state as $\rho_0=\rho(0)$ and denote the final state after encoding with controls and noises as $\rho_x=\rho(T)$. Normally, the standard deviation of estimating $x$ can be quantified by the quantum Cram{\'e}r-Rao bound, i.e., $\Delta x \geq 1/ \sqrt{F_Q}$ \cite{GI11}, where $F_Q$ represents the quantum Fisher information (QFI). For a general final state $\rho_x$, its QFI can be calculated by \cite{PS18}
\begin{equation}\label{QFI}
F_Q(\rho_x)=\sum_{p,q;\lambda_p+\lambda_q>0} \frac{2}{\lambda_p+\lambda_q}|\langle p| \partial_x \rho_x|q\rangle|^2,
\end{equation}
where $p,q$ and $\lambda_p,\lambda_q$ are the eigenvalues and eigenvectors of $\rho_x$. Furthermore, in consideration of the cost of  encoding time $T$, we can introduce the sensitivity $\upsilon$ to more  carefully evaluate the metrology performance, i.e., 
\begin{equation}
\upsilon=\frac{\sqrt{T}}{\gamma_c \sqrt{F_Q }}\label{sensitivity},
\end{equation}
where $\gamma_c$ is the transduction parameter, for example, the gyromagnetic ratio for estimating magnetic fields.

\subsection{Optimization of the controls}
To achieve the best metrology performance, the key task is optimizing the adjustable controls. For convenience, we divide the total encoding time $T$ into $K=T/\Delta t$ slices.
 Thus, the noiseless Hamiltonian becomes $\mathcal{H}[k]=\mathcal{H}_0(x)+\sum_{l=1}^L u_l[k] \mathcal{H}_l$, with $k=1,2,...,K$. The final system state can then be calculated by 
\begin{equation}\label{dynamics-control}
|\rho(T)\rangle= \Pi_{k=1}^K e^{\mathcal{\hat L}[k]\Delta t}|\rho(0)\rangle =\Pi_{k=1}^K  \varepsilon_{x,C}^k (\Delta t)|\rho(0)\rangle,	
\end{equation} 
where $\varepsilon_{x,C}^k$ denotes the $k$th sliced dynamical evolution superoperator. Now the problem becomes finding the optimal sliced control fields $\boldsymbol{u}=(u_l[k])$, with $k=1,2,...,K$ and $l=1,2,...,L$, to maximize the QFI or minimize the sensitivity.  Many optimization algorithms can be used to accomplish this task; here, we choose the Nelder-Mead simplex algorithm \cite{nelder1965simplex}.
It is a multidimensional unconstrained direct-search algorithm without resorting to gradients. Due to its simplicity, it has been successfully implemented in various advanced quantum control experiments \cite{kelly2014optimal,egger2014adaptive,frank2017autonomous}. 
It is based on applying some geometric transformations, including reflection, expansion, contraction, and shrinkage, on an initialized working simplex which consists of many vertices representing the to-be-optimized parameter vectors. The vertices are renewed in the optimal-solution direction iteratively until the stopping criterion is met.  
In our optimization, we use the MATLAB function FMINSEARCH to conveniently include this method.

\section{Numerical Simulations}\label{numerical simulations }
To demonstrate the effectiveness of the proposed control-enhanced scheme,  we apply it to the problem of frequency estimation under some typical Markovian noise channels, including parallel dephasing, transverse dephasing, and amplitude damping. Specifically, we consider estimating the frequency $\omega_0$ along $z$ axis using one or two carbon nuclear spins. By comparing the metrology metrics QFI and sensitivity, we show the advantage of our control-enhanced scheme over the standard and ancilla-assisted schemes.

\subsection{Parallel-dephasing channel}

\begin{figure}
	\centering
	\includegraphics[width=0.50\textwidth,height=0.50\textwidth]{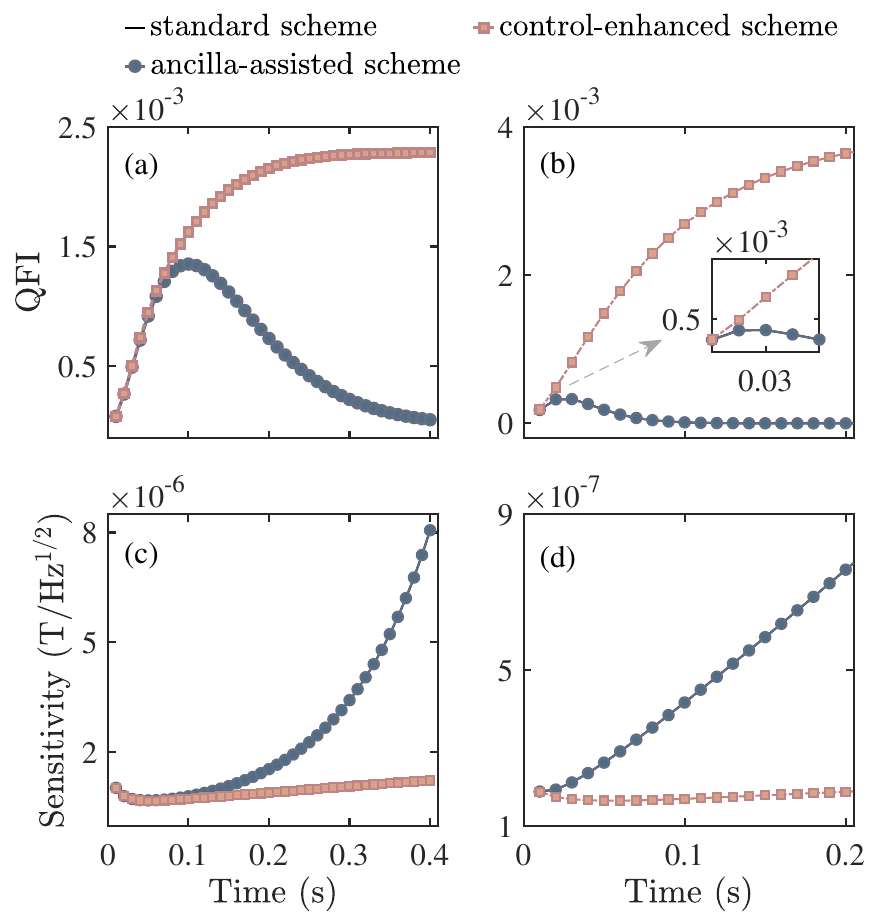}
	\caption{Numerical comparison of three quantum metrology schemes on frequency estimation under parallel-dephasing noise. (a) and (c) show the QFI and the sensitivity vs the encoding time for single-qubit dephasing noise, where we set $\gamma=1/T_2=10~\text{s}^{-1},$ and $\omega_0=2\pi$.  (b) and (d) plot the QFI and the sensitivity vs the encoding time under two-qubit uncorrelated dephasing noise, where we set $\gamma^n=1/T_2^n=10~\text{s}^{-1},$ with $n=1,2$ and $\omega_0=2\pi$. 
	In all the plots, the curve of the standard scheme and that of the ancilla-assisted scheme coincide with each other.
}  
	\label{dephasing}
\end{figure}

Parallel dephasing is a common dominant noise source for many physical platforms \cite{childress2006coherent,dutt2007quantum}.
For the parallel-dephasing channel, the corresponding Lindblad operators are $L_1=\sigma_z/\sqrt{2}$ and $L_2=\mathbb{I}$; thus, the Lindblad equation in Eq. (\ref{linda}) can be explicitly written as 
\begin{equation}\label{single-par}
\frac{d\rho}{dt}=-i[\mathcal{H},\rho]+\frac{\gamma}{2}(\sigma_z\rho\sigma_z-\rho),
\end{equation}
% thus the above  Lindblad operator are $L_1=\sigma_z/\sqrt{2}$ and $L_1=\sigma_z/\sqrt{2}$
where $\gamma=1/T_2$ and $T_2$ characterizes the coherence time. We consider transverse controls; thus, the encoding Hamiltonian in Eq. (\ref{ham}) becomes $\mathcal{H}(\omega_0)=\omega_0\sigma_z/2 + u_x(t)\sigma_x/2+u_y(t)\sigma_y/2$, with $x=\omega_0$.
Without loss of generality, we set $\gamma=10~\text{s}^{-1}, \omega_0=2\pi$ in our simulations; see the results in Figs. \ref{dephasing}(a) and \ref{dephasing}(c). From Fig. \ref{dephasing}(a), it can be seen that when the encoding time $T$ is smaller than the coherence time $T_2$, neither the ancilla-assisted scheme nor our control-enhanced scheme can visibly improve the QFI compared to the standard scheme. 
The reason is that the decoherence effect has little influence on the metrology performance within the coherence time; thus, the standard scheme is already optimal.
However, when increasing the encoding time beyond $T_2$, the QFI for the standard scheme quickly decays because of the severe decoherence. The ancilla-assisted scheme also fails to increase the QFI due to the fact that maximally entangled states will lose their advantages compared to the uncorrelated states in the presence of decoherence \cite{HM97}. Remarkably, our control-enhanced scheme can maintain a QFI increment for the encoding time that is far beyond $T_2$; thus, the sensitivity can be improved by almost an order of magnitude compared to the standard scheme and the ancilla-assisted scheme, as shown in Fig. \ref{dephasing}(c).

The above results demonstrate the effectiveness of our control-enhanced scheme for improving the metrology performance under one-qubit parallel-dephasing noise; we now explore its abilities in the two-qubit case.
 For simplicity, we assume that the two qubits are uncorrelated; thus, the Lindblad operators can be expressed as $L_n=\sigma_z^n/\sqrt{2},n=1,2$, and the corresponding Lindblad equation in Eq. (\ref{linda}) can be rewritten as   
\begin{equation}
	\frac{d\rho}{dt}
	=-i[\mathcal{H},\rho]+ \sum_{n=1}^2\frac{\gamma_n}{2}(\sigma_{z}^n \rho\sigma_{z}^n-\rho),
\end{equation}
with 
\begin{equation}
\mathcal{H}=\sum_{n=1}^2 \left[\frac{1}{2} \omega_0 \sigma_z^n+ u_x^n(t) \sigma_x^n/2+ u_y^n(t) \sigma_y^n/2 \right],
\end{equation}
where $\sigma_\alpha^n (\alpha=x,y,z; n=1,2)$ represent the Pauli operators for the $n$th qubit, $u_x^n$ and $u_y^n$ are the transverse controls applied to the $n$th qubit, and $\gamma_{n}=1/T_2^{n}$ characterize the corresponding dephasing rates.
 We set $\gamma_1=\gamma_2=10~\text{s}^{-1}$ and $\omega_0=2\pi$ in our simulations; see the results in Figs. \ref{dephasing}(b) and \ref{dephasing}(d).
 From Fig. \ref{dephasing}(b), it can be seen that the optimal encoding time using the standard scheme is much smaller than $T_2^1(T_2^2)$;  this is because the multi-qubit dephasing noises quickly destroy the system coherence, thus achieving an even smaller QFI than the single-qubit case in Fig. \ref{dephasing}(a). Similarly, we find that  the ancilla-assisted scheme cannot improve the metrology performance compared to the standard scheme.  
 By implementing  our control-enhanced scheme, the QFI can be greatly increased, and the sensitivity in Fig.  \ref{dephasing}(d) can also be enhanced, up to around an order of magnitude. 
This verifies the effectiveness of our method in the two-qubit parallel-dephasing noise environment.

\subsection{Transverse-dephasing channel}

\begin{figure}
	\centering
	\includegraphics[width=0.50\textwidth,height=0.50\textwidth]{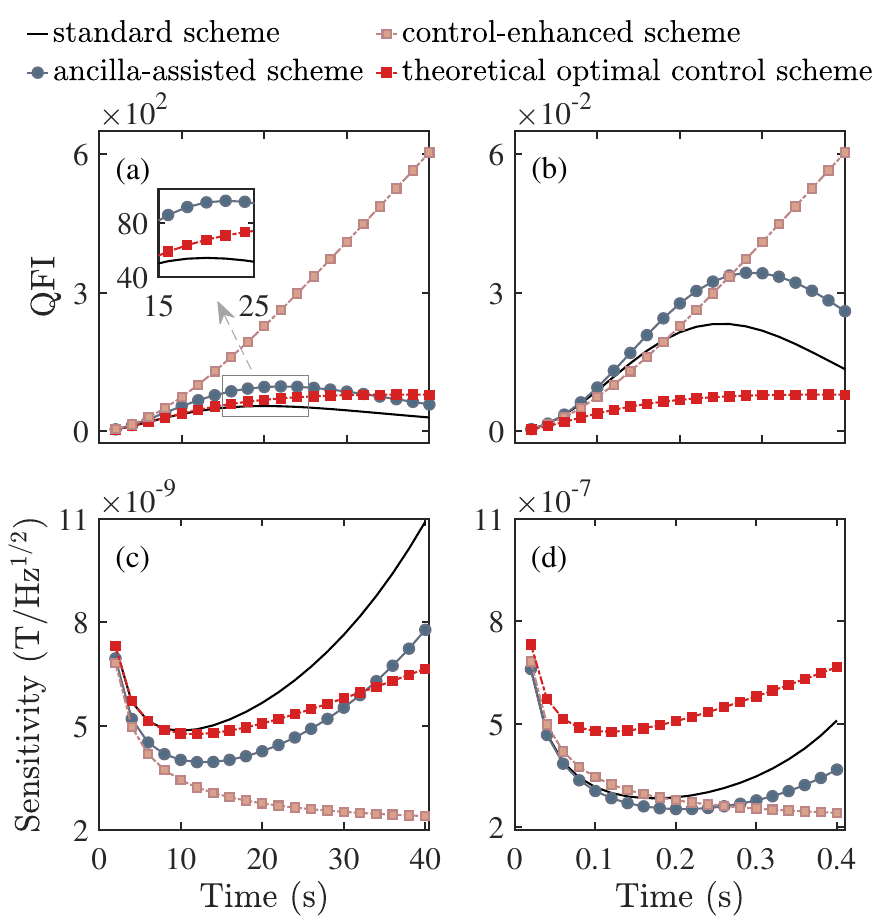}
	\caption{Numerical comparison of four quantum metrology schemes on frequency estimation under transverse-dephasing noise. (a) and (c) show the QFI and the sensitivity vs the encoding time for $\gamma=0.1~\text{s}^{-1}$ and $\omega_0=2\pi$. (b) and (d) demonstrate similar cases, but with  $\gamma=10~\text{s}^{-1}$.}	
	\label{transverse_dephasing}
\end{figure}

For transverse-dephasing noise, the Lindblad operators are $L_1=\sigma_x/\sqrt{2}$ and $L_2=\mathbb{I}$; thus, the corresponding Lindblad equation in Eq. (\ref{linda}) can be explicitly expressed as 
\begin{equation}
\frac{d\rho}{dt}=-i[\mathcal{H},\rho]+\frac{\gamma}{2}(\sigma_x\rho\sigma_x-\rho),
\end{equation}
where $\gamma$ denotes the transverse-dephasing rate. Here, we consider longitudinal controls; thus, the encoding Hamiltonian in Eq. (\ref{ham}) becomes $\mathcal{H}(\omega_0)=\omega_0\sigma_z/2 + u_z(t)\sigma_z/2$, with $x=\omega_0$. In our simulations, we set $\omega_0=2\pi$ and consider two cases with the distinct dephasing rates $\gamma=0.1~\text{s}^{-1}$ and $\gamma=10~\text{s}^{-1}$. In addition to the standard scheme, the ancilla-assisted scheme, and our control-enhanced scheme, we demonstrate the results using  the theoretical optimal controls predicted in Ref. \cite{LJY17}, namely, $u_x(t)=u_y(t)=0 $ and $u_z(t)=-\omega_0$; this is called the theoretical optimal control scheme.

For a small dephasing rate $\gamma=0.1~\text{s}^{-1}$, the simulation results are shown in Figs. \ref{transverse_dephasing}(a) and \ref{transverse_dephasing}(c).
From Fig. \ref{transverse_dephasing}(a), it can be seen that the optimal encoding time with the standard scheme satisfies $T_{\rm opt} \backsimeq 2/\gamma=20~\text{s}$ (see the inset), which is consistent with the conclusion in Ref. \cite{LJY17} under the  condition $\omega_0 \gg \gamma$. 
When the encoding time is sufficiently small  ($T<4~\text{s}$),  we find that all the schemes have similar performances, which is due to the same reason as in the previous parallel-dephasing case; that is, the decoherence effect in a short time is so small that additional controls are useless for improving the metrology precision.  
However, as the encoding time increases, the four schemes gradually show different features. 
Specifically, when the encoding time satisfies $4~\text{s}< T< 20~\text{s}$, it can be seen that  all three advanced schemes can moderately increase the QFI compared to the standard scheme. This phenomenon is different from that in the parallel-dephasing case. Herein, the transverse-dephasing noise is perpendicular to the encoding operator, which makes  the additional controls easier to manipulate to resist the noises.
When the encoding time $T> 20~\text{s}$, the QFI of the theoretical optimal control scheme and the ancilla-assisted scheme slowly increase or begin to decrease, while the QFI of our control-enhanced scheme can still significantly grow.  From the perspective of sensitivity, our scheme can achieve a two-fold to three-fold improvement compared to the other three schemes, as shown in Fig. \ref{transverse_dephasing}(c).

The above results demonstrate the effectiveness of our control-enhanced scheme under transverse-dephasing noise with a  relatively small dephasing rate, we now explore the case with a large dephasing rate, namely,  $\gamma=10~\text{s}^{-1}$ [see Figs. \ref{transverse_dephasing}(b) and \ref{transverse_dephasing}(d)]. In this case, the optimal encoding time $T_{\rm opt}$ using the standard scheme is obviously larger than $2/\gamma$. The theoretical optimal control scheme cannot improve the QFI compared to the standard scheme during the entire tested encoding time.  The noiseless  Hamiltonian in Eq. (\ref{ham}) equals zero with the theoretical optimal controls; thus, the system dynamics is totally determined by the transverse-dephasing noise. Such significant dephasing effects make the performance of the theoretical optimal scheme quickly decrease, unlike in the previous case with  the small dephasing rate. The ancilla-assisted scheme  can improve the QFI by around 50\% compared to the standard scheme but gets very little sensitivity improvement, as shown in Fig. \ref{transverse_dephasing}(d). Our control-enhanced scheme improves the QFI up to three-fold compared to the standard scheme. Similarly, it achieves only a relatively small sensitivity improvement [see Fig. \ref{transverse_dephasing}(d)].  These results show that the transverse-dephasing noise with large dephasing rates is relatively harder to resist when trying to improve the metrology.

%Overall, the above results demonstrate the effectiveness of our control-enhanced scheme on improving the metrology performance under transverse dephasing noise, with either small or large dephasing rate. 

 \subsection{Amplitude-damping channel}
  \begin{figure}
 	\centering
 	\includegraphics[width=0.50\textwidth,height=0.30\textwidth]{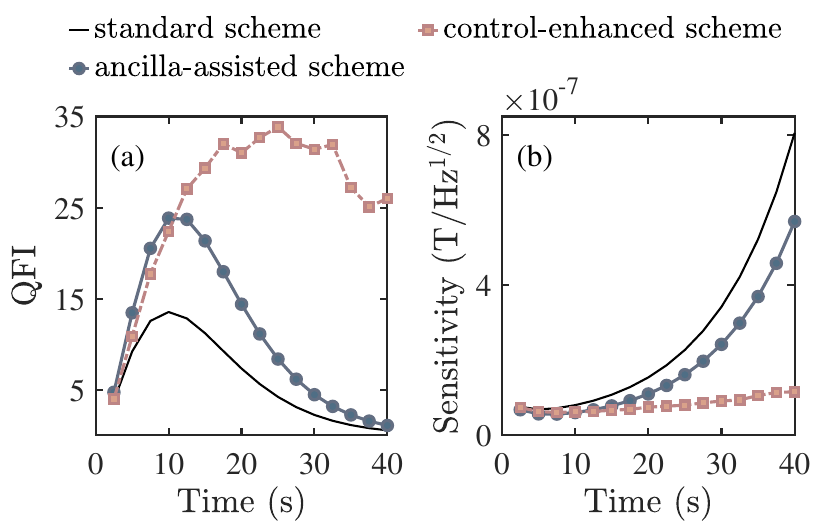}
 	\caption{Numerical comparison of three quantum metrology schemes on frequency estimation under amplitude-damping noise. (a) shows the QFI vs the encoding time. (b) plots the corresponding sensitivity vs the encoding time. In the simulations, we set $\gamma=2/T_1=0.2~\text{s}^{-1}$ and $\omega_0=2\pi$.}
 	\label{amplitude damping}
 \end{figure}
The generalized amplitude-damping channel characterizes the effect of dissipation at non-zero temperature, and dominates the noise for many physical systems \cite{fujiwara2004estimation,chirolli2008decoherence}. 
The Lindblad operators of this channel are $L_1=\sigma_{-}$ and $L_2=\sigma_{+}$, where $\sigma_{\pm}=(\sigma_x \pm i \sigma_y)/2$; thus, the Lindblad equation in Eq. (\ref{linda}) can be written as 
\begin{equation}\label{Spontaneous}
 \begin{aligned}
 	\frac{d\rho}{dt}
 	&=-i[\mathcal{H},\rho]+\gamma_{-}\left [\sigma_{-}\rho\sigma_{+}-\frac{1}{2}\{ \sigma_{+}\sigma_{-},\rho\} \right]\\
 &+\gamma_{+} \left[\sigma_{+}\rho\sigma_{-}-\frac{1}{2}\{ \sigma_{-}\sigma_{+},\rho\} \right],
 \end{aligned}
\end{equation}
where $\gamma_{\pm}$ represents the amplitude-damping rates. 
Considering transverse controls,  the encoding Hamiltonian in Eq. (\ref{ham}) turns out to be $\mathcal{H}(\omega_0)=\omega_0\sigma_z/2 + u_x(t)\sigma_x/2+u_y(t)\sigma_y/2$, with $x=\omega_0$.
For  simplicity,  we  set $\gamma_{+}=0$ and denote $\gamma_{-}=\gamma$.
 Suppose that the environment is at zero temperature; then the generalized amplitude-damping channel reduces to the amplitude-damping  channel \cite{chirolli2008decoherence}.
In this case,  we have $\gamma=1/T_1$, where $T_1$ usually characterizes the spin-lattice relaxation time.
Without loss of generality, we set $\gamma=0.2~\text{s}^{-1}$ and $\omega_0=2\pi$ in our simulations; see the results in Figs. \ref{amplitude damping}(a) and \ref{amplitude damping}(b).

From Fig. \ref{amplitude damping}(a), we find that the optimal encoding time in the standard scheme satisfies the predicted relation $T_{\rm opt} = 2/\gamma=2 T_1=10~\text{s}$ \cite{LJY17}.
When the encoding time is smaller than $T_{\rm opt}$, it is clear that the ancilla-assisted scheme and our control-enhanced scheme have comparable performance; both can achieve great enhancement of the QFI compared to the standard scheme.
However, when the encoding time is beyond $T_{\rm opt}$, we observe that the QFI using the ancilla-assisted scheme starts to decay,  while our control-enhanced scheme still maintains the increment of the QFI.
If we investigate the sensitivity, as shown in  Fig. \ref{amplitude damping}(b), we find that the ancilla-assisted scheme improves the sensitivity by only around 25$\%$, while our control-enhanced scheme leads to almost an order of magnitude improvement compared to the standard scheme.

\section{Experiment}\label{experiment}
 \begin{figure*}
 	\centering
 	\includegraphics[width=0.90\textwidth,height=0.44\textwidth]{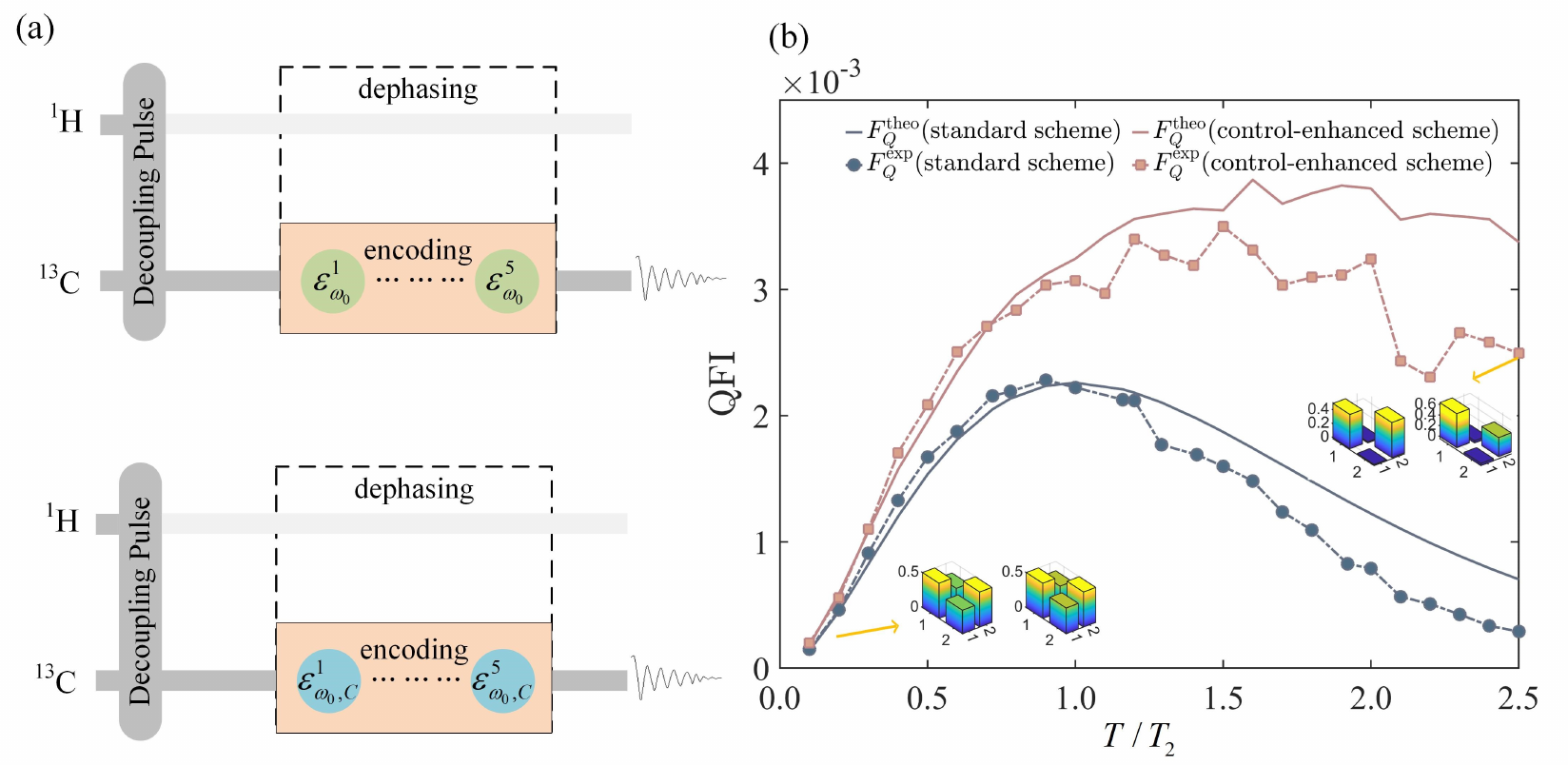}
 	\caption{Experimental comparison of the standard scheme and our control-enhanced scheme on frequency estimation in the NMR system. (a) The top and the bottom panels show the schematic diagrams for the standard scheme and the control-enhanced scheme, respectively. The nuclear spin $^{1}\text{H}$ is decoupled, and the spin $^{13}\text{C}$ is initialized at $|+\rangle$ for the standard scheme and a random state for the control-enhanced scheme. The encoding process is realized by freely evolving the system with an offset $\omega_0$ in the presence of pure parallel-dephasing noise, marked as $\varepsilon_{\omega_0}^k,k=1,2,...,5	$. We set $\omega=60\times 2\pi$, and the measured coherence times is  $T_2=0.149~\text{s}$.
 	In our control-enhanced scheme, the encoding dynamics is engineered by additional control for resisting the noises, marked as $\varepsilon_{\omega_0,C},k=1,2,...,5$. The final state $\rho_{\omega_0}$  and its perturbed state $\rho_{\omega_0+\delta \omega_0}$, with $\delta \omega=2\pi$, are measured from experiments for calculating the QFI.
 	 	(b) shows the theoretically calculated QFI $F_Q^{\text{theo}}$ and the experimentally measured QFI $F_Q^{\text{exp}}$ for the standard scheme and the control-enhanced scheme. We also demonstrate the tomography results of the initial states and the final states (exact states and perturbed states) using our control-enhanced scheme.}
 	\label{exp}
 \end{figure*}

To verify the effectiveness of our control-enhanced quantum  metrology scheme, we test it in the nuclear-magnetic-resonance (NMR) system. We use  $^{13}\text{C}$-labeled chloroform dissolved in acetone-d6 to perform experiments on a Bruker Avance III 400-MHz spectrometer at room temperature. Specifically, we decouple the $^1\text{H}$ nuclear spin and take the $^{13}\text{C}$ nuclear spin as a probe for estimating the frequency $\omega_0$ along the $z$ axis with transverse controls. The experimental schematic diagrams are shown in Fig. \ref{exp}(a), where we compare the standard scheme and our control-enhanced scheme.  In the following, we describe the experimental procedures and demonstrate the corresponding experimental results.

\subsection{Experimental procedures}
We first introduce the way to calculate the evolution dynamics to predict the system's final state. 
As parallel dephasing is the dominant noise in the NMR system \cite{VL05}, we utilize the Lindblad equation in Eq. (\ref{single-par}) to model the encoding dynamics, i.e., $\frac{d\rho}{dt}=-i[\mathcal{H},\rho]+\frac{\gamma}{2}(\sigma_z\rho\sigma_z-\rho)$, with $\mathcal{H}(\omega_0)=\omega_0\sigma_z/2 + u_x(t)\sigma_x/2+u_y(t)\sigma_y/2$. Now the key task becomes obtaining the dephasing rate $\gamma=1/T_2$. A practical and reasonable method is to first obtain the spectrum of the thermal equilibrium state, then measure the width-at-half-height of the spectrum $\Gamma$, and finally get the value of $T_2$ using the relation $T_2=1/(\pi\Gamma)$ \cite{levitt2013spin}. 

Next, we introduce how to experimentally measure the QFI for quantifying the metrology performance. In practice, we use the following equation related to Uhlmann’s quantum fidelity \cite{cerezo2021sub} to equivalently measure the QFI instead of  Eq. (\ref{QFI}): 
\begin{equation}\label{sub-qfi}
F_Q(\rho_{\omega_0})=8\lim_{\delta \omega_0 \rightarrow 0}\frac{1-\text{Tr}\sqrt{\sqrt{\rho_{\omega_0}}\rho_{\omega_0+\delta \omega_0} \sqrt{\rho_{\omega_0}} }}{\delta^2 \omega_0},
\end{equation}
where $\delta \omega_0$ is a small change in $\omega_0$ and $\rho_{\omega_0}$ and $\rho_{\omega_0+\delta \omega_0}$ are the exact state and the perturbed state, respectively. It is worth noting that we need to choose appropriate $\delta \omega_0$ so that the difference between $\rho_{\omega_0}$ and $\rho_{\omega_0+\delta \omega_0}$ can  easily be distinguished experimentally and the estimation of the QFI is accurate enough.

Based on the above strategies, we now describe how to perform the metrology tasks using the standard scheme and the control-enhanced scheme [see Fig. \ref{exp}(a)]. For the standard scheme, the $^{1}\text{H}$ nuclear spin is decoupled, and the nuclear spin $^{13}\text{C}$ is prepared in the state $|
    +\rangle= (|0\rangle +|1\rangle)/\sqrt{2}$.
    The system is then encoded by freely evolving with an offset $\omega_0$ along the $z$ axis. This process automatically includes the dephasing effects of the NMR system. Afterwards, the system's final state and its  perturbed state are measured to calculate the corresponding QFI. 
    For our control-enhanced scheme, the initial state of the $^{13}\text{C}$ nuclear spin is randomly prepared. The encoding process is also realized by freely evolving with an offset $\omega_0$ along the $z$ axis. However, we add additional transverse controls in this process. 
     Specifically, we divide the total encoding time $T$ into $K=5$ slices, and the encoding dynamics with controls can be solved by Eq.  (\ref{dynamics-control}). Thus, in each time period $\Delta t= T/K$,  three kinds of effects exist, including the encoding, the controls, and the dephasing. We observe the system's final state and its  perturbed state to calculate the corresponding QFI. To search optimal controls for maximizing the QFI, we use the ``FMINSEARCH''  function in MATLAB, as described above.

\subsection{Experimental results}
The measured result of the width-at-half-height of the spectrum is $\Gamma=2.13$ Hz; thus, $T_2=1/\pi \Gamma \approx 0.149$ s. After a careful test, we choose  $\omega_0=60 \times 2\pi$ and $\delta \omega_0=2\pi$ in our experiments. We compare the standard scheme and the control-enhanced scheme for a total encoding time up to $2.5T_2$, as shown in Fig. \ref{exp}(b). We label the QFI directly calculated by the numerical results as $F_Q^{\text{theo}}$ and the QFI measured from experiments as $F_Q^{\text{exp}}$.
 When the encoding time is smaller than the coherence time ($T/T_2<1$), it can be seen that  $F_Q^{\text{theo}}$ and $F_Q^{\text{exp}}$ match very well. This reveals that within the coherence time, our theoretical model is accurate, and our control-enhanced scheme has the same performance as predicted.
 However, as the encoding time increases ($T/T_2>1$), the gaps between $F_Q^{\text{theo}}$and $F_Q^{\text{exp}}$ gradually become distinct. This may result from many factors, such as the amplitude-damping effects or other unknown noises. As our theoretical model  concerns only the parallel-dephasing noise, many other kinds of noises can lead to a reduction of the theoretically predicted QFI. 
 Nevertheless, during the tested encoding time, we find that our control-enhanced scheme can improve the QFI compared to the standard scheme, up to around twofold. 
 Overall, the above experimental results are roughly consistent with the theoretical predictions, revealing the effectiveness of the proposed control-enhanced quantum metrology scheme.

\section{Conclusion and discussion}\label{conclusion}
In this work, we proposed a practical control-enhanced quantum metrology scheme to defend against Markovian noises to improve the  precision of frequency estimation. 
The numerical comparisons of our scheme with the standard scheme and the ancilla-assisted scheme  under typical noise channels revealed that our scheme works for all the tested cases and can achieve substantial precision improvements up to an order of magnitude. The experimental verification in the nuclear-magnetic-resonance system shows the effectiveness of the proposed method.

Instead of using gradient-based algorithms or reinforcement learning algorithms, we utilized a gradient-free algorithm that requires fewer computational resources and is more friendly for experimental applications.
Therefore, our scheme can be easily adapted to a fully online version for automatically discovering optimal controls on real platforms \cite{PhysRevLett.123.040501}.  As realistic noises may be very complex and other unknown imperfections may be involved, our scheme should achieve more impressive improvements \cite{PhysRevA.102.062605,yang2020probe}. 
Moreover, the research here can be combined with  related theoretical studies \cite{FA17,ED11,KJ13} which provide the precision bound of noisy quantum metrology to explore the underlying properties of the encoding dynamics in our scheme. 
In addition, our scheme can also be extended to quantum metrology in a non-Markovian noise environment \cite{PhysRevLett.102.090401,PhysRevLett.107.130404}.

\section*{Acknowledgments} 
This work was supported by the National Natural Science Foundation of China (Grants No. 12204230, No. 1212200199, No. 11975117, No. 12075110, No. 11905099, No. 11875159, No. 11905111, No. U1801661, and No. 92065111); National Key Research and Development Program of China (Grant No. 2019YFA0308100); Guangdong Basic and Applied Basic Research Foundation (Grants No. 2019A1515011383 and No. 2021B1515020070); Guangdong Provincial Key Laboratory (Grant No. 2019B121203002); Guangdong International Collaboration Program (Grant No. 2020A0505100001); Shenzhen Science and Technology Program (Grants No. RCYX20200714114522109 and No. KQTD20200820113010023); Science, Technology, and Innovation Commission of Shenzhen Municipality (Grants No. ZDSYS20190902092905285, No. KQTD20190929173815000, No. JCYJ20200109140803865, and No. JCYJ20180302174036418); and Pengcheng Scholars, Guangdong Innovative and Entrepreneurial Research Team Program (Grant No. 2019ZT08C044).

%\bibliography{Reference}
%apsrev4-2.bst 2019-01-14 (MD) hand-edited version of apsrev4-1.bst
%Control: key (0)
%Control: author (8) initials jnrlst
%Control: editor formatted (1) identically to author
%Control: production of article title (0) allowed
%Control: page (0) single
%Control: year (1) truncated
%Control: production of eprint (0) enabled
\providecommand{\noopsort}[1]{}\providecommand{\singleletter}[1]{#1}%

\end{document}